\documentclass[prl,superscriptaddress,twocolumn,showpacs]{revtex4}
\usepackage{amsmath,amssymb,graphicx,color,bm,soul}

\begin{document}

\title{Optomechanics with Cavity Polaritons: Dissipative Coupling and Unconventional Bistability}

\author{O. Kyriienko}\email{kyriienko@ukr.net}
\affiliation{Science Institute, University of Iceland, Dunhagi-3, IS-107, Reykjavik, Iceland}
\affiliation{Division of Physics and Applied Physics, Nanyang Technological University 637371, Singapore}

\author{T. C. H. Liew}
\affiliation{Division of Physics and Applied Physics, Nanyang Technological University 637371, Singapore}

\author{I. A. Shelykh}
\affiliation{Science Institute, University of Iceland, Dunhagi-3, IS-107, Reykjavik, Iceland}
\affiliation{Division of Physics and Applied Physics, Nanyang Technological University 637371, Singapore}

\begin{abstract}
We study a hybrid system formed from an optomechanical resonator and a cavity mode strongly coupled to an excitonic transition inside a quantum well. We show that due to the mixing of cavity photon and exciton states, the emergent quasiparticles --- polaritons --- possess coupling to the mechanical mode of both dispersive and dissipative nature. We calculate the occupancies of polariton modes and reveal bistable behavior, which deviates from conventional Kerr nonlinearity or dispersive coupling cases due to the dissipative coupling. The described system serves as a good candidate for future polaritonic devices and solid state quantum information processing.
\end{abstract}
\date{\today}

\pacs{71.36.+c, 42.50.Wk, 07.10.Cm, 42.65.Pc, 71.35.-y}


\maketitle

{\it Introduction.---} \emph{Cavity optomechanics} is a field of physics which studies hybrid systems of optical resonators coupled to mechanical oscillators \cite{Reviews,AspelmeyerRev}. A central role there is played by phenomena of radiative pressure and dynamical backaction, which allow optical control of the mechanical system. An ultimate milestone of cavity optomechanics is  optomechanical cooling of the mechanical resonator leading to achievement of the long-thought regime where physics on the boundary of classical and quantum mechanics can be studied. Being realized recently in various optomechanical systems \cite{First}, quantum optomechanical coupling triggered numerous proposals and experimental observations, of both applied and fundamental interest, including quantum non-demolition measurements \cite{OConnell}, possible achievement of the standard quantum limit \cite{ClerkRev,Anetsberger,Teufel,Purdy}, protocols for quantum computing \cite{Qcomp}, quantum communication \cite{Lukin} and optomechanical entanglement \cite{Entanglement}, optical bistability \cite{Dorsel}, strong optomechanical coupling \cite{Marquardt2007,Groblacher2009}, optomechanically induced transparency \cite{Weis,Safavi-Naeini}, photon blockade and single-photon emission \cite{Rabl2011,Xu2013}, tests of quantum gravity \cite{Pikovski}, and many others. Moreover, interest in the field tends to grow \cite{AspelmeyerRev}.

Cavity optomechanics is essentially a hybrid area of physics, largely involving other components and subsystems to increase the number of applications. In this fashion, optomechanical systems with coupling to single atoms \cite{Hammerer2009b}, collective spins \cite{CollSpin}, superconducting qubits \cite{OConnell}, cold atom Bose-Einstein condensates (BECs) \cite{Zhang2010}, quantum dots \cite{Rundquist2011,Hughes2013} and carbon nanotubes \cite{Imamoglu2012} were studied. Furthermore, the experimental unification of solid-state physics with cavity optomechanics became possible with the growth of semiconductor structures first in a vibrating disk \cite{Ding} or membrane \cite{Usami2012} geometry, and recently in a conventional VCSEL structure \cite{Fainstein2013}. Notably, the expansion of optomechanics to the multimode case revealed strong nonlinearities in given systems and was proven to be useful for the generation of non-classical states and quantum information processing \cite{AspelmeyerRev}.

The aforementioned optomechanical systems are based on the conventional \emph{dispersive} coupling mechanism, which originates from the mechanical modulation of the cavity photon frequency. Recently it was realized that another type of photon-phonon coupling, namely \emph{dissipative} coupling, is possible \cite{Elste2009}. It appears due to the mechanical modulation of the cavity photon damping rate, and was shown to allow an optomechanical cooling in the bad cavity limit \cite{Elste2009}, generate reactive optical force \cite{Li2009} and squeezing \cite{Huang2010}, and lead to Fano line shapes in the force spectrum \cite{Weiss2013}. However, other optomechanic effects modified with a dissipative coupling are yet to be studied.

Another widely studied branch of non-linear optics, which originates from the strong light-matter coupling between microcavity photons and excitonic transitions, is \emph{polaritonics} \cite{CarusottoRev}. The resulting mixed quasiparticles --- exciton-polaritons --- obey bosonic statistics, have very small effective mass ($\sim 10^{-5}$ of free electron mass) and can form non-equilibrium BECs at relatively high temperatures \cite{Kasprzak2006,Baumberg2008}. The Kerr non-linearity in a polariton medium, appearing from the exciton-exciton interaction, enables polariton bistability \cite{Baas2004}, which was shown to be extremely useful for realization of optical circuits \cite{Liew2008,Sanvitto2011,Ballarini2013} and optical memory \cite{Cerna2013}.

In this Letter we propose to merge the physics of optomechanics and polaritonics, with both phonon-photon and strong exciton-photon coupling being present. This crucially changes the nature of optomechanical coupling leading to simultaneous dispersive and dissipative coupling of the phonon mode to the polariton state. Using the master equation we calculate the steady state solutions for polariton occupation numbers and analyze the bistable behavior coming from two mechanisms of optomechanical coupling. Additionally, we point out that an emergent squeezer-type Hamiltonian for the phonon subsystem appears. The described findings can be used for future solid state quantum information applications.

\begin{figure}
\centering
\includegraphics[width=0.9\linewidth]{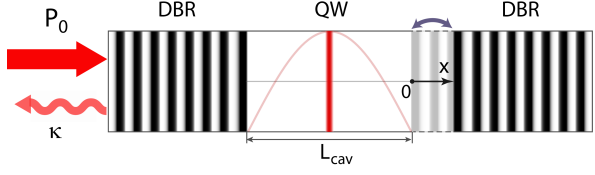}
\caption{(color online) Sketch of the system. Optical and mechanical resonators formed by two distributed Bragg reflectors (DBR) with a QW placed in the antinode of an optical cavity. Black (white) stripes correspond to GaAs (AlAs) layers. We consider the case of mechanical modulation with a positive displacement $x$ which decreases the cavity frequency. The point $x=0$ corresponds to maximal frequency of the cavity mode, $\omega_{cav}$. Here $P_0$ denotes the pumping rate of an external laser, and $\kappa$ refers to the decay rate of the cavity photon.}
\label{fig:Fig1}
\end{figure}

{\it The model.---} We study an optomechanical resonator formed by a micropillar with moveable Bragg reflectors, recently realized experimentally in Ref. \cite{Fainstein2013}, supplemented with an undoped quantum well (QW) placed in the antinode of the resonator [Fig. \ref{fig:Fig1}]. Considering a sample with comparably high mirror quality factor ($Q_{opt} \sim 10^4$), the strong coupling between cavity photons and two-dimensional QW excitons is possible. The generic Hamiltonian of the system can be written as:
\begin{align}
\notag
\hat{\cal{H}}=&\hbar \omega_{cav}(x) \hat{a}^{\dagger} \hat{a} + \hbar \omega_{exc} \hat{c}^{\dagger} \hat{c} + \frac{\hbar \Omega}{2} (\hat{a}^{\dagger} \hat{c} + \hat{c}^{\dagger} \hat{a}) + \hbar \Omega_{m} \hat{b}^{\dagger} \hat{b}\\
&+ \hbar P_{0} e^{-i \omega_{p} t} \hat{a}^{\dagger} + \hbar P_{0} e^{i \omega_{p} t} \hat{a},
\label{eq:H_in}
\end{align}
where $\hat{a}$, $\hat{b}$ and $\hat{c}$ correspond to field operators for cavity photons, phonons (oscillation quanta of the mirrors) and QW excitons, respectively, with bare energies $\hbar \omega_{cav}(x)$, $\hbar \Omega_{m}$ and $\hbar \omega_{exc}$. The presence of the mechanical resonator makes the $\hbar \omega_{cav}(x)$ displacement dependent. The third term corresponds to exciton-photon coupling, with $\hbar \Omega$ being the Rabi splitting between polariton modes. Finally, the last two terms in the Hamiltonian (\ref{eq:H_in}) describe the continuous wave (cw) coherent pumping of the cavity mode, at rate $P_{0}$ and frequency $\omega_{p}$. For brevity, in Hamiltonian (\ref{eq:H_in}) we neglect exciton-exciton interactions, targeting optomechanical effects.

The strong exciton-photon coupling causes hybridization of the modes and the first three terms in Hamiltonian (\ref{eq:H_in}) can be diagonalized using lower polariton (LP) and upper polariton (UP) eigenstates \cite{SM}. To allow a simple quantum description of the system we consider the case of small exciton-photon detuning, $\delta = \omega_{cav} - \omega_{exc} \ll \Omega$, and the typical case of small optomechanical displacement, where a linear expansion of the cavity mode frequency can be made, $\omega_{cav}(x)\approx \omega_{cav} + x \partial \omega_{cav}/\partial x + \mathcal{O}[x]^{2}$ \cite{AspelmeyerRev}. Using a quantum description, the displacement operator reads $\hat{x} = x_{ZPF} (\hat{b} + \hat{b}^{\dagger})$, where $x_{ZPF}$ is a zero-point fluctuation amplitude. The LP and UP states are separated in energy and decoupled, such that one can neglect the UP state under low frequency excitation, which has limited occupation due to thermalization processes.

We now divide the Hamiltonian (\ref{eq:H_in}) into four terms:
\begin{equation}
\hat{\mathcal{H}}=\hat{\mathcal{H}}_{0}+\hat{\mathcal{H}}^{(1)}_{int}+\hat{\mathcal{H}}^{(2)}_{int} + \hat{\mathcal{H}}_{p}.
\label{eq:H}
\end{equation}
$\hat{\mathcal{H}}_{0}$ refers to the energy of free LP and phonon modes,
\begin{align}
\hat{\mathcal{H}}_{0}=\hbar \widetilde{\omega}_{L} \hat{a}_{L}^{\dagger} \hat{a}_{L} + \hbar \Omega_{m} \hat{b}^{\dagger} \hat{b},
\label{eq:H0}
\end{align}
where the modified LP frequency is
\begin{align}
\label{eq:omega_L_s}
\widetilde{\omega}_{L} = \frac{\omega_{cav}+\omega_{exc}}{2} - \frac{\Omega}{2}\big(1 + \frac{\delta^{2}}{2 \Omega^{2}} \big) - \frac{g_{0}^2}{4 \Omega^2},
\end{align}
and $g_0$ denotes the vacuum optomechanical strength \cite{AspelmeyerRev}.

The interaction part of the Hamiltonian comprises two terms. The first term describes the conventional dispersive polariton-phonon coupling,
\begin{align}
\hat{\mathcal{H}}^{(1)}_{int}=-\frac{\hbar g_{0}}{2} \hat{a}^{\dagger}_{L} \hat{a}_{L}(1-\delta / \Omega) (\hat{b}^{\dagger} + \hat{b}),
\label{eq:Hint_1}
\end{align}
where a weak nonlinear polariton-phonon coupling can be controlled by the exciton-photon detuning $\delta$.

The second interaction term can be written as
\begin{align}
\hat{\mathcal{H}}^{(2)}_{int}=-\frac{\hbar^2 g_{0}^2}{4\hbar \Omega} \hat{a}^{\dagger}_{L} \hat{a}_{L} (\hat{b}^{\dagger} \hat{b}^{\dagger} + \hat{b} \hat{b}) -\frac{\hbar^2 g_{0}^2}{2\hbar \Omega} \hat{a}^{\dagger}_{L} \hat{b}^{\dagger} \hat{a}_{L} \hat{b}.
\label{eq:Hint_2}
\end{align}
The first term is quadratic in the phonon operators, corresponding to the typical squeezer Hamiltonian, which can lead to the appearance of non-classical phonon states suitable for quantum computational schemes. The second term describes polariton-photon scattering.

Finally, the pumping term for LPs contains both purely static and mechanically modulated terms:
\begin{align}
\label{eq:Hp}
\hat{\mathcal{H}}_{p}= &-\frac{\hbar P_{0}}{\sqrt{2}} \Big(1 - \frac{\delta}{2\Omega} \Big) (e^{-i \omega_{p} t} \hat{a}^{\dagger}_L + e^{i \omega_{p} t} \hat{a}_L) \\ \notag &- \frac{\hbar P_{0}}{2 \sqrt{2}}\frac{g_{0}}{\Omega} (\hat{b}^{\dagger}+ \hat{b}) (e^{-i \omega_{p} t} \hat{a}^{\dagger}_L + e^{i \omega_{p} t} \hat{a}_L).
\end{align}

So far we have rewritten the coherent part of the Hamiltonian and have shown that strong exciton-photon coupling leads to modification of the optomechanical coupling. Now let us consider the changes that it introduces in the incoherent part of dynamics. This can be treated using the master equation approach \cite{Carmichael} for the density matrix $\rho$ of the system where $\dot{\rho}=-i/\hbar \big[ \hat{\mathcal{H}}, \rho \big] + \hat{\mathcal{L}}\rho$, and $\hat{\mathcal{L}}\rho$ corresponds to the Lindblad superoperator, which describes decay of the cavity photon and exciton modes. Equivalently, we can write the decay terms using the polariton picture \cite{SM}. Finally, for the relevant case of the semiclassical approximation, the traced value of the LP Lindblad superoperator reads
\begin{align}
\notag \mathrm{Tr} \lbrace \hat{\mathcal{L}}\rho[\hat{a}_{L}] \rbrace = &\mathrm{Tr} \lbrace \Big[ \widetilde{\kappa}_{L} + \frac{g_{0}}{\Omega} \frac{(\kappa - \kappa_{exc})}{2} \big(\hat{b}^{\dagger} + \hat{b} \big) \Big] \Big( \hat{a}_{L} \rho \hat{a}_{L}^{\dagger} - \\ &-\hat{a}_{L}^{\dagger} \hat{a}_{L} \rho /2- \rho \hat{a}_{L}^{\dagger} \hat{a}_{L}/2 \Big) \rbrace,
\label{eq:Lindblad_L}
\end{align}
where we introduced the LP decay rate, $\widetilde{\kappa}_{L} = (\kappa + \kappa_{exc})/2 - \delta (\kappa - \kappa_{exc})/2 \Omega$,
%
%
with $\kappa$ and $\kappa_{exc}$ being decay rates of bare cavity photons and excitons, respectively.

One can see that the decay rate for polaritons is modified, since it shares contributions from both cavity photons and excitons. Moreover, the second term in the first brackets is influenced by the mechanical system. This corresponds to \emph{dissipative} coupling between phonons and polaritons. It linearly depends on the ratio of phonon-photon to exciton-photon interaction constants, as well as the difference of decay rates of the modes, and can be modified for particular semiconductor structures. Similarly, due to relations between pump and decay, the same considerations are valid for the coherent pumping term (\ref{eq:Hp}), where an analog of the dissipative coupling appears.

\textit{Equations of motion.---} Knowing the coherent Hamiltonian written for LP states and their mechanically modulated dissipation, we proceed to derive dynamic equations for the mean values of the operators $\hat{a}_{L}$ and $\hat{b}$. Here we will focus on the quasiclassical regime, which is judicious for studying optical bistability. Also, we disregard the interaction term $\hat{\mathcal{H}}^{(2)}_{int}$ in the Hamiltonian (\ref{eq:Hint_2}), which is quadratic in the phonon-photon coupling constant.

Using the master equation, we can write the dynamic equations for the mean value of, \textit{e. g.}, the LP field:
\begin{align}
\dot{\bar{a}}_{L}= -\frac{i}{\hbar} \langle \big[ \hat{a}_{L}, \hat{\mathcal{H}} \big] \rangle + {\rm Tr} \{ \hat{a}_{L} \hat{\mathcal{L}}\rho [\hat{a}_{L}]\},
\label{eq:aL_eom_in}
\end{align}
where $\bar{a}_{L} \equiv \langle\hat{a}_{L} \rangle = {\rm Tr} \{ \rho \hat{a}_{L} \}$. The commutators in Eq. (\ref{eq:aL_eom_in}) can be evaluated using bosonic commutation relations, the trace term rearranged using its cyclic properties, and in the lowest order mean-field approximation $\langle \hat{a}_{i} \hat{a}_{j} .. \hat{a}_{k} \rangle \approx \bar{a}_{i} \bar{a}_{j} .. \bar{a}_{k}$. This mean field approximation is fully justifiable for the case of relatively high occupation numbers, assuming strong pumping of cavity mode.

Repeating this procedure for the phonon field, we obtain a closed system of quasiclassical equations:
\begin{align}
\notag
\dot{\bar{a}}_{L}&= i\Delta_{L} \bar{a}_{L}  + i g_{0} \left(1- \frac{\delta}{\Omega} \right) \underline{\mathrm{A}} {\rm Re}\{ \bar{b} \} \bar{a}_{L} + \frac{i P_0}{\sqrt{2}} \left( 1 - \frac{\delta}{2\Omega} \right) \\ \label{eq:aL_dot} &+ \frac{i P_0}{\sqrt{2}}\frac{g_0}{\Omega} \underline{\mathrm{B}} {\rm Re}\{ \bar{b} \} - \frac{\widetilde{\kappa}_{L}}{2}\bar{a}_{L} - (\kappa - \kappa_{exc}) \frac{g_0}{\Omega} \underline{\mathrm{B}} {\rm Re} \{ \bar{b} \} \bar{a}_{L},\\
\notag
\dot{\bar{b}}= &-i\Omega_{m}\bar{b} + \frac{ig_0}{2}\left( 1- \frac{\delta}{\Omega} \right)\underline{\mathrm{A}}|\bar{a}_{L}|^{2} + \frac{i P_0}{\sqrt{2}} \frac{g_0}{\Omega} \underline{\mathrm{B}} \mathrm{Re} \{ \bar{a}_{L} \} \\  &- \Gamma_m \bar{b} /2,
\label{eq:b_dot}
\end{align}
where we used the frame rotating at the pump frequency, with $\Delta_L=\omega_p - \widetilde{\omega}_L$ being the laser-LP detuning, and introduced a decay of the mechanical oscillations with rate $\Gamma_{m}$. The second terms in the RHS of Eqs. (\ref{eq:aL_dot})--(\ref{eq:b_dot}) correspond to typical dispersive couplings, which are modified due to strong exciton-photon coupling. For the LP field evolution Eq. (\ref{eq:aL_dot}), the dissipative coupling appears in both pump and decay terms, being the fourth and the last terms, respectively. On the contrary, the dissipative coupling for the phonon dynamics enters only in the third term in Eq. (\ref{eq:b_dot}) corresponding to the pump. For convenience we introduced the constants $\underline{\mathrm{A}}$ and $\underline{\mathrm{B}}$, which take values $0$ and $1$, and allow for switching between the dispersive and the dissipative coupling cases.

Finally, given the dynamic equations, we find steady-state solutions of the system (\ref{eq:aL_dot})--(\ref{eq:b_dot}) setting $\dot{\bar{a}}_{L}=0$ and $\dot{\bar{b}}=0$. Additionally, we can study the stability of these solutions analysing the spectrum of fluctuations \cite{SM,Sarchi2008}.

\textit{Results and discussion.---} Let us now set realistic parameters, considering a system similar to that studied in Ref. \cite{Fainstein2013}. For a GaAs$/$AlAs $\lambda /2$ micropillar cavity, the cavity wavelength is equal to $\lambda = 870$ nm and we consider a cavity mode lifetime of $\tau=1/\kappa=5$ ps. The mechanical resonator is characterized by frequency $\Omega_{m}/2\pi=20$ GHz, lifetime $\tau_{m}=\hbar/\Gamma_{m}=60$ ns,  $x_{ZPF}=5.8\times 10^{-7}$ nm, and vacuum optomechanical strength $g_{0}/2\pi=4.8\times 10^7$ Hz. The exciton-photon coupling for a single GaAs QW typically gives a Rabi splitting $\Omega/2 \pi = 0.48$ THz, which can be increased using a larger number of QWs. The exciton lifetime is estimated as $\tau_{exc}=1/\kappa_{exc}=0.5$ ns, and the variable exciton-cavity detuning $\delta$ will be used to control the nature of optomechanical coupling.
\begin{figure}[t]
\centering
\includegraphics[width=1.0\linewidth]{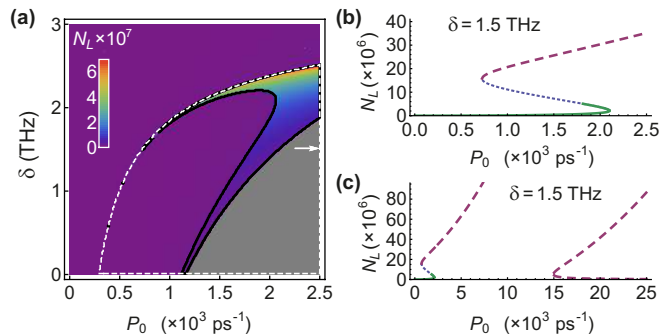}
\caption{(color online) (a) Phase diagram showing the LP occupation of the highest intensity stable state for varying $\delta$ and $P_0$. The dotted white curve represents the boundary between stable and unstable solutions, with the gray region being parametrically unstable. The solid black curve indicates the region where two stable states exist and the system is bistable. The white arrow indicates a cross-section for which power dependences are shown in (b)-(c) for low (b) and high (c) pump intensity. (b) Dispersively driven S-shape mean field solution for low pump intensity, showing a zoomed region of plot (c). (c) Large scale behavior revealing appearance of dissipatively driven solutions. In (b)-(c) the green solid curves represent stable states, blue dotted curves show the saddle node instability, while magenta dashed curves represent parametrically unstable states.}
\label{fig:Fig2}
\end{figure}

First, we consider the case of negative laser detuning $\Delta_{L}=-1.5$ THz [Fig. \ref{fig:Fig2}], which is typically required for multibranch solutions in the case of dispersive coupling \cite{AspelmeyerRev}. In Fig. \ref{fig:Fig2}(a) we show the phase diagram of the system as a function of pump intensity $P_0$ and exciton-photon detuning $\delta$. It reveals a large region with parametrically unstable solutions (white dashed area), which corresponds to the presence of Hopf bifurcation in the system, while a bistable region is denoted by the black curves. The former can be explained by the fact that the chosen parameters correspond to the unresolved sideband regime with small mechanical damping, where the optomechanical system deviates from the Kerr-nonlinear-like behavior \cite{Aldana2013}. The LP occupation numbers $N_L=|\bar{a}_L|^2$ are given in Fig. \ref{fig:Fig2}(b,c) for positive exciton-photon detuning $\delta = 1.5$ THz, showing low (b) and high (c) pump intensity regions. In the low pump strength region we observe an S-shaped behaviour of the LP occupation number, characteristic of dispersive coupling. Here the bistable window is present, accomplished by the single-mode unstable middle branch, and parametrically unstable upper branch [Fig. \ref{fig:Fig2}(b)]. At higher pump intensity, a dissipative coupling mechanism comes into play, leading to the appearance of a second branch solution [Fig. \ref{fig:Fig2}(c), right]. However, for the chosen system parameters it is parametrically unstable.
\begin{figure}[t]
\centering
\includegraphics[width=1.0\linewidth]{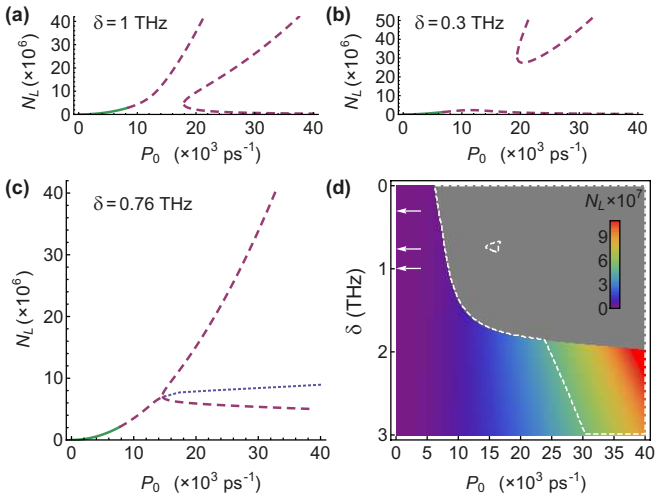}
\caption{color online) (a, b, c): Stationary solutions plotted for positive laser detuning $\Delta_L = 4.25$ THz, and exciton-photon detuning $\delta = 1.,~0.3,~0.76$ THz, respectively. (d): Phase diagram showing stability of solutions, with white arrows corresponding to the detunings $\delta$ chosen in (a)-(c). The notation coincides with that of Fig. \ref{fig:Fig2}.}
\label{fig:Fig3}
\end{figure}

Next, we examine the case of positive laser detuning $\Delta_L$, which is usually overlooked in the dispersive coupling case, being characterized by the single mode optical limiter solution. However, here the dissipative coupling plays a major role, leading to the aforementioned double-branch solution for large exciton-photon detuning $\delta$ [Fig. \ref{fig:Fig3}(a)], and a modified unstable branch for small detuning [Fig. \ref{fig:Fig3}(b)]. An intriguing behavior of the modes is revealed for the case of intermediate detuning $\delta$ shown in Fig. \ref{fig:Fig3}(c), where three solutions incorporate two branches with Hopf bifurcation corresponding to unstable behavior and a middle single mode unstable branch, though much different from conventional S-shape form. We verify that this behavior is a result of complex interplay between both dispersive and dissipative couplings by switching on and off the couplings, finding that for $\underline{\mathrm{A}}=0,~\underline{\mathrm{B}}=1$ and $\underline{\mathrm{A}}=1,~\underline{\mathrm{B}}=0$ only single solutions are present~\cite{SM}. In Fig. \ref{fig:Fig3}(d) we supplement the mean-field solutions with a phase diagram, indicating the stable and parametrically unstable regions. Numerical modelling of the system dynamics reveals self-sustained oscillations in the mechanical amplitude $x=x_{ZPF}(\bar{b}+\bar{b}^*)$, representing a potential tunable saser~\cite{SM}. Additionally we observed anharmonic oscillations of polariton density, leading to Q-switching behavior of a polariton laser~\cite{SM}.
\begin{figure}[t]
\centering
\includegraphics[width=1.0\linewidth]{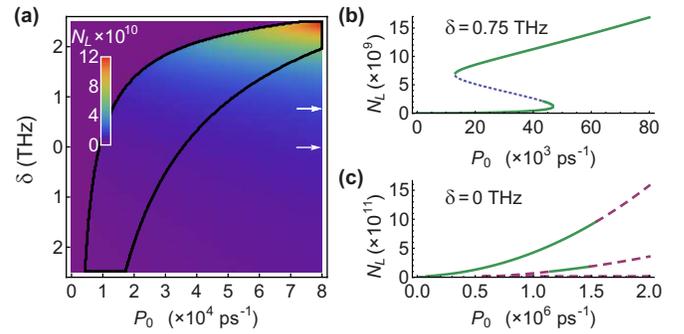}
\caption{(color online) Unconventional bistability of the modified system, where mechanical oscillator damping is increased, $\tau_{m}=1.2~10^{-14}$ s, $Q_m =2$, and negative laser detuning $\Delta_{L}=-1.5$ THz. (a) Phase diagram showing a large bistable window, and absence of parametric instability. (b) Stationary solutionts with S-shaped bistability present at small pump intensity. (c) Solutions for high pump intensity showing unconventional bistability, fully driven by the dissipative coupling.}
\label{fig:Fig4}
\end{figure}

Finally, in Fig. \ref{fig:Fig4} we present calculations for a theorized system, where parameters satisfy the resolved sideband regime with large mechanical damping, where $\Gamma_m > \kappa$, and the mechanical resonator quality factor $Q_m$ is of the order of unity. This allows to enlarge greatly the region of stable solutions [Fig. \ref{fig:Fig4}(a)]. Here a bistable region appears for the S-shaped solution at small pump intensities [Fig. \ref{fig:Fig4}(b)]. Moreover, it reappears in the high pumping region, which is fully governed by the dissipative coupling mechanism, manifesting an unconventional bistability present in the system.

\textit{Conclusions.---} We considered an optomechanical system, where a cavity mode is additionally strongly coupled to a quantum well exciton. Due to the modification of the eigenstates of the system, the mechanical coupling contains both dispersive and dissipative channels. This strongly modifies the stationary states of the system, which can demonstrate both unconventional bistable and parametrically unstable behavior. The results are important for future optical circuits.\\

The work was supported by FP7 IRSES projects POLAPHEN and POLATER, and FP7 ITN NOTEDEV network. O.K. acknowledges the support from the Eimskip fund.

\end{document}